\documentclass[conference]{IEEEtran}
\usepackage{verbatim}
\usepackage[utf8x]{inputenc}
\usepackage{epstopdf}
\usepackage[pdftex]{graphicx}
\usepackage{graphics}
\usepackage{tabularx}
\usepackage{booktabs}
\usepackage[table]{xcolor}
\usepackage{multirow}
\usepackage{siunitx}
\usepackage{tabu}
\usepackage{booktabs} \usepackage{ctable}
\usepackage[cmex10]{amsmath}
\usepackage{amssymb}
\usepackage{cite}
\usepackage{textcomp}
\usepackage{array}
\newcolumntype{C}[1]{>{\centering\let\newline\\\arraybackslash\hspace{0pt}}m{#1}}

\graphicspath{{./Figures/}}

\ifCLASSOPTIONcompsoc
\usepackage[tight,normalsize,sf,SF]{subfigure}
\else
\usepackage[tight,footnotesize]{subfigure}
\fi

\begin{document}
\title{Multiband Spectrum Sensing: Challenges and Limitations} 

\author{\IEEEauthorblockN{Ghaith Hattab}
\IEEEauthorblockA{ECE, Queen's University\\
Kingston, ON, Canada\\
Email: g.hattab@queensu.ca}
\and
\IEEEauthorblockN{Mohammed Ibnkahla}
\IEEEauthorblockA{ECE, Queen's University\\
Kingston, ON, Canada\\
Email: mohamed.ibnkahla@queensu.ca}}

\maketitle

\begin{abstract}
Multiband spectrum access presents the next generation of cognitive radio networks (CRNs), where multiple bands are sensed and accessed to enhance the network's throughput, improve spectrum's maintenance, and reduce handoff frequency and data transmission interruptions due to the activities of the primary users. In this paper, we discuss the challenges and limitations of the major multiband spectrum sensing techniques. Particularly, we highlight the edge-detection problem and examine several issues of the state-of-the-art wavelet-based techniques. We also study the compressive sensing problem. Finally, we highlight the promises of utilizing the angle-domain for the CRNs.
\end{abstract}


\section{Introduction}
\IEEEPARstart{M}{ultiband} spectrum access has recently become a major area of research due to the promising enhancements it brings to cognitive radio networks (CRNs) \cite{Hattab2,Ibnkahla2,Sun2}. The basic principle is to enable the secondary users (SUs) to simultaneously access multiple bands. In one hand, this will tangibly improve the network's throughput since by accessing multiple bands, more spectral opportunities will be utilized. On the other hand, When SUs access multiple bands, then data transmission interruptions due to the sudden reappearance of primary users (PUs) become less since SUs can seamlessly handoff from one band to another. In fact, the handoff frequency could be reduced if proper channel allocation schemes are utilized. For instance, the SU could stop transmission over the channels that are only being reclaimed by the PU and remain transmitting over the rest of the accessed channels.

Multiband spectrum access is enabled by multiband spectrum sensing techniques. Tremendous research has been done for single-band sensing techniques, and even though this constitutes the core of the multiband sensing, more effort is still required towards realizing these techniques for multiband spectrum access. In general, the single-band detectors are categorized into three main types: The energy detector, the coherent detector, and the feature detector \cite{Axell1}. The energy detector is the simplest technique since the SU does not require prior knowledge of the PU signal, yet it has a poor performance in the low signal-to-noise ratio (SNR) regime \cite{Sahai1}. The coherent detector maximizes the SNR (at the output of the SU detector), but it requires a complete knowledge of the PU signal, and this is generally rendered infeasible in practice \cite{Cabric1}. The feature detector exploits some signal features (e.g., pilots) to improve detection at low SNR at the expense of additional signal processing complexity \cite{Hattab3}.

The multiband spectrum sensing techniques can be categorized into three broad categories \cite{Hattab2}. The first one is based on serial sensing, which is essentially a single-band detector that sequentially senses multiple bands, one at a time \cite{Joshi}. This technique is relatively slow, and it requires tunable filters and oscillators, in which tuning and sweeping must be done carefully. The second category is based on parallel sensing, which is enabled by equipping the SU's receiver with multiple single-band detectors. The detectors operate on parallel such that each one senses a single band \cite{Quan1}. Clearly, processing multiple channels in parallel provides faster sensing compared to serial sensing. Nevertheless, the receiver structure is bigger, more complex, and more expensive because multiple detectors must be implemented on the same device. The third category is based on wideband  sensing, which is the focus of this paper. This category includes wideband-specific detectors such as the wavelet-based detectors, the compressive sensing-based detectors, etc.

In this paper, we define the multiband detection problem and discuss various wideband-based spectrum sensing techniques. In particular, we present the edge-detection problem and study the challenges and limitations of the recent wavelet-based techniques. We also present the compressive sensing problem and discuss the main issues that may hinder its implementation in the future. In addition, we highlight the potentials of the angle-domain, which can be utilized for multiband sensing to provide additional spectral opportunities.

The rest of the paper is organized as follows. The multiband detection problem is defined in Section II. Wavelet-based and compressive-based sensing are presented in Section III and Section IV, respectively. Section V highlights the potentials of the angle-domain for multiband sensing. Finally, the conclusions are summarized in Section VI.

\section{The Multiband Spectrum Sensing Problem}

\begin{figure}[!b]
\centering
\includegraphics[width=3.5in]{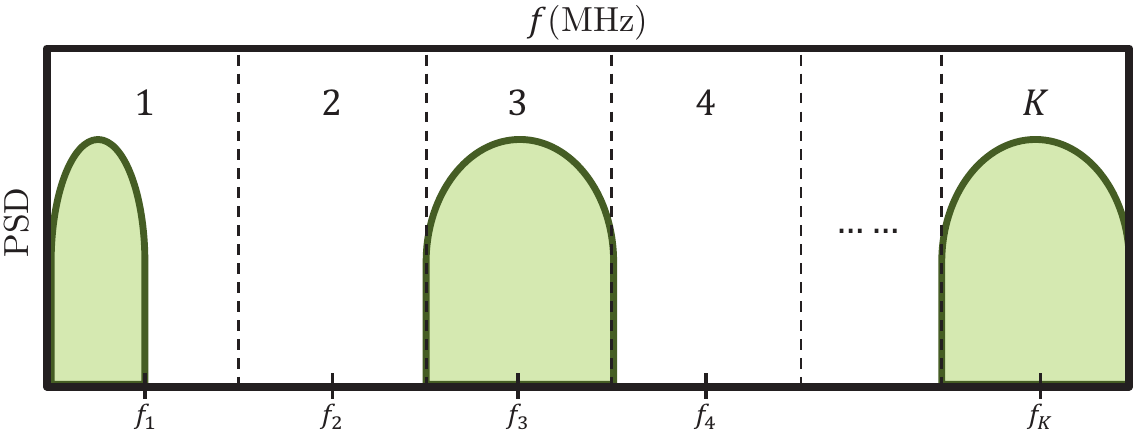}
\caption{A wideband spectrum divided into non-overlapping subchannels.}
\label{fig:WidebandSpectrum}
\end{figure}

In multiband spectrum sensing, the SU must sense a wide spectrum to exploit more spectral opportunities. This spectrum can be divided into multiple subchannels. Therefore, the multiband sensing problem is essentially sensing multiple subchannels. For example, Fig. \ref{fig:WidebandSpectrum} shows the power spectral density (PSD) of a wideband spectrum that is divided into $K$ subchannels. It is commonly assumed, in the literature, that these subchannels are not overlapping and have identical bandwidths \cite{Quan1,Beaulieu3}.

The SU's task is to decide which subchannels are available and which ones are occupied by other users. The nature of the wideband spectrum makes multiband sensing a difficult task. The reasons are as follows:
\begin{itemize}
\item The available bands, in general, are not necessarily contiguous. For instance, in Fig. \ref{fig:WidebandSpectrum}, the 2nd and the 4th subchannels are unoccupied, whereas the 3rd subchannel is occupied. Therefore, the SU must be able to properly configure its transmission to avoid interference with the PUs (e.g., use orthogonal frequency division multiplexing (OFDM) technology as a physical layer \cite{Jun3}).
\item Some wireless devices consume small portion of the bandwidth, yet the entire bandwidth may be considered unavailable. For example, in IEEE 802.22, wireless microphones consume a portion of a 6MHz channel, yet the channel will be considered occupied \cite{GwangZeen}. Thus, the 1st subchannel in Fig. \ref{fig:WidebandSpectrum} must not be utilized by the SU.
\item In practical communications, the occupancy of the subchannels are correlated. An example of correlated subchannels is a PU signal transmitting over multiple bands such as the PU signals in Wireless LAN (WLAN) or broadcast television \cite{Hossain}. A more challenging issue is if a PU transmits a wideband signal, and a portion of its bandwidth is in a deep fade.  If the SU is not aware of the correlation of the subchannels, it will consider the deep-faded portion as an unoccupied subchannel, and hence the SU interferes with the PU when the former accesses the band.
\end{itemize}

If the occupancy of the subchannels is independent, then the multiband spectrum sensing problem is formulated as
\begin{equation}
\label{eq:MultibandDetectionProblem}
\begin{aligned}
\mathcal{H}_0^k:  &~  \mathbf{r}_k= \mathbf{w}_k                 \\
\mathcal{H}_1^k:  &~  \mathbf{r}_k= \mathbf{x}_k+\mathbf{w}_k,
\end{aligned}
\end{equation}
where $k=1,2,\ldots,K$ is the subchannel index, $\mathbf{r}_k=[r^1_k,r^2_k,\ldots,r^N_k]^T$ is the received signal at the SU's receiver, $\mathbf{x}_k$ is the transmitted PU signal, and $\mathbf{w}_k$ is noise.  The SU decides $\mathcal{H}_0^k$ if the $k$-th subchannel is unoccupied, and it decides $\mathcal{H}_1^k$ if it is occupied. In other words, the SU solves $K$ binary hypothesis testing problems. Thus, the complexity of this problem linearly increases as the number of subchannels increases. On the contrary, if the subchannels are correlated, then the complexity of solving the detection problem grows exponentially with $K$ \cite{Hossain}.

The decision rule for each subchannel is basically the likelihood ratio test, which has the following basic form
\begin{equation}
\label{eq:MBDecisionRule}
T(\mathbf{r}_k)    \overset{\mathcal{H}_1^k}{\underset{\mathcal{H}_0^k}{\gtrless}} \xi_k,
\end{equation}
where $T(\mathbf{r}_k)$ is the test statistic of the $k$-th subchannel, and $\xi_k$ is the threshold that divides the decision region into $\mathcal{H}_1^k$ and $\mathcal{H}_0^k$. For instance, one of the most common parallel sensing techniques is to implement multiple energy detectors in parallel (e.g., the multiband joint detector \cite{Quan1}). Thus, the test statistic of the $k$-th energy detector is
\begin{equation}
E_k\triangleq\sum_{m=0}^{N_F-1}|R_k(m)|^2   \overset{\mathcal{H}_1^k}{\underset{\mathcal{H}_0^k}{\gtrless}} \xi_k,
\end{equation}
where $R_k(m)$ is the fast Fourier transform (FFT) of $\mathbf{r}_k$, and $N_F$ is the FFT size. The authors in \cite{Quan1} have demonstrated that significant throughput gains can be accomplished if the thresholds of the $K$ subchannels are jointly optimized to achieve a predetermined detection performance.

\section{Wavelet-Based Spectrum Sensing}

\subsection{The Edge Detection Problem}
In the previous model, two assumptions are made about the available knowledge at the SU side: the number of subchannels, $K$, and their corresponding center frequencies $\{f_k\}$. However, in practice, the SU may not have prior knowledge of how many subchannels are there in the wideband spectrum (since each PU network can have different requirements for channel bandwidth, carrier frequency, etc.). In that case, the SU may need to get this knowledge, and thus the \emph{edge detection problem} can be used as an alternative system model.

Wavelet transform is a powerful tool to detect singularities \cite{Tian1}. In the CR context, these singularities are observed at the boundaries (edges) of the subchannels. Thus, by detecting these singularities, the SU can estimate the boundaries of these subchannels, and hence locate their center frequencies. After that, the SU can estimate the occupancy of the spectrum within these boundaries.

The continuous wavelet transform (CWT) can be carried out in the frequency-domain or in the time-domain. For instance, in the frequency-domain, the CWT is
\begin{equation}
\label{eq:WaveletTransformFrequency}
\mathcal{W}_s(f)   =   \hat{R}(f) * \psi_s(f),
\end{equation}
where $*$ is the convolution operator, $\hat{R}(f)$ is the wideband PSD as a function of frequency, and $\psi(f)$ is the \emph{wavelet smoothing function}. This function is dilated by a factor $s$ such that $\psi_s(f)   =   (1/s)\psi(f/s)$ (typically, $s=2^j$, for $j=1,2,\ldots,J$).

The PSD of the wideband spectrum has irregularities at the edges of the subchannels. By taking the derivatives of the CWT, the edges can be further sharpened. Mallat and Hwang have shown in \cite{Mallat} that the edges of the subchannels correspond to the local maxima of the first derivatives. This technique is referred as the \emph{wavelet modulus maxima} (WMM). However, due to the characteristics of the wideband spectrum, noise may impact detecting the edges, and to overcome this, a product of several first derivatives of the CWT is implemented to suppress noise and help sharpen the edges \cite{Tian1}. This technique is referred as the \emph{wavelet multiscale product} (WMP), and it is expressed as
\begin{equation}
\label{eq:MultiscaleProductWavelet}
\mathcal{P}_J   = \prod_{j=1}^{J} \mathcal{W}'_{s=2^j}(f),
\end{equation}
where $\mathcal{W}'_s(f)$ is the first derivative of $\mathcal{W}_s(f)$. Increasing $J$ provides a better performance at the expense of higher complexity.

\subsection{Challenges}
\begin{figure}[!b]
\centering
\includegraphics[width=3.5in]{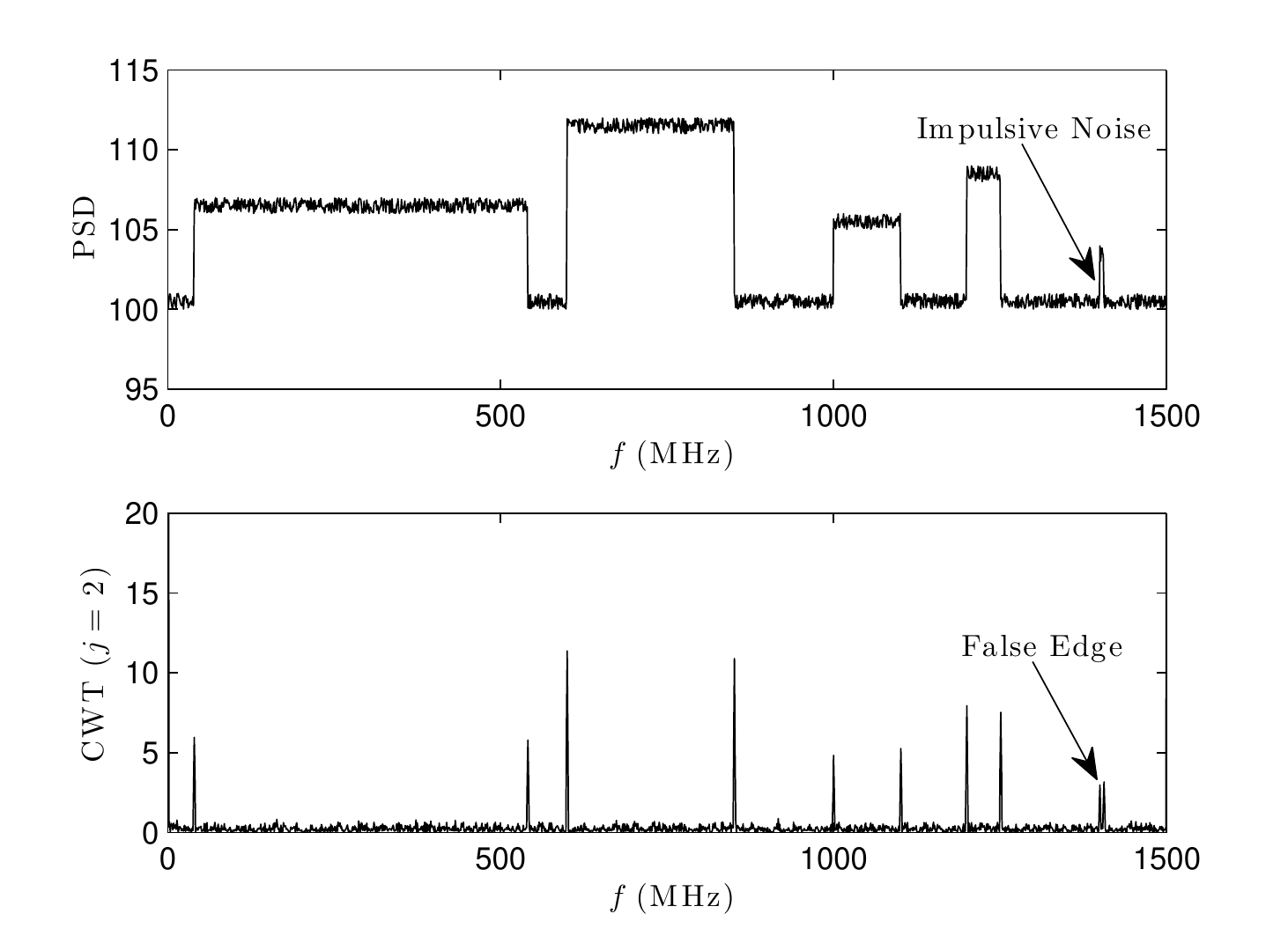}
\caption{The wideband spectrum and the edge estimation using the CWT ($j=2$).}
\label{fig:CWT}
\end{figure}

While wavelet-based spectrum sensing techniques stand out in edge-detection, there are several challenges that limit or degrade their performance. We have assumed so far that the spectrum is smooth within the subchannel and changes abruptly when a transition occur from one channel to another. This is not always true due to the impulsive noise, which poses a challenge when estimating the edges' location. Thus, the wavelet-based techniques may falsely consider the noise as a subchannel edge. For instance, consider Fig. \ref{fig:CWT}, where the PSD of a wideband spectrum and the CWT is shown. Here, we use a first-order Daubechies wavelet. Due to the presence of the impulsive noise at $f=1400$MHz, the SU will falsely detect an edge there.

Ignoring false-edges is vital, and one of the techniques to correctly reject them is to use a threshold, $\eta$, such that if a local maxima of (\ref{eq:MultiscaleProductWavelet}) is less than $\eta$, then it is rejected \cite{Zeng2}. Since the local maximum points depend on the amplitude of their corresponding PSD, $\eta$ may vary widely. To overcome this problem, Zeng et al. suggest normalizing the WMP using the mean of the PSD \cite{Zeng2}. Thus, (\ref{eq:MultiscaleProductWavelet}) becomes
\begin{equation}
\label{eq:MultiscaleProductWaveletNormalized}
\mathcal{\hat{P}}_{J}   = \frac{1}{(\bar{E})^J}\prod_{j=1}^{J} \mathcal{W}'_{s=2^j}(f),
\end{equation}
where $\bar{E}=\frac{1}{K}\sum_{k=1}^{K} E_k$. We remark that it is a challenging task to set the appropriate $\eta$ \cite{Zeng2}. Particularly, now it is possible that a heavily noise-corrupted true-edge is misdetected when its WMP is less than $\eta$.

Another drawback of the WMP is that narrowband signals with slow variations cannot be accurately detected because using the product attenuates their corresponding edges. To mitigate this, the WMP can be replaced with the \emph{wavelet multiscale sum} (WMS) \cite{Xu}. Mathematically, this is expressed as
\begin{equation}
\label{eq:MultiscaleSumWavelet}
\mathcal{S}_J   = \sum_{j=1}^{J} \mathcal{W}'_{s=2^j}(f).
\end{equation}
Another key issue is the type of the wavelet function. In particular, it is shown in \cite{Xu} that using orthogonal-based smoothing functions is troublesome when $J$ increases in the WMS technique. In contrary, the edge-detection is possible at different variations of the scale $J$ when \emph{non-orthogonal} smoothing functions (e.g., Gaussian wavelet family) are implemented. However, non-orthogonal family provides a poor performance when the SNR is very low  \cite{Xu}.

\begin{figure}[!t]
\centering
\includegraphics[width=3.5in]{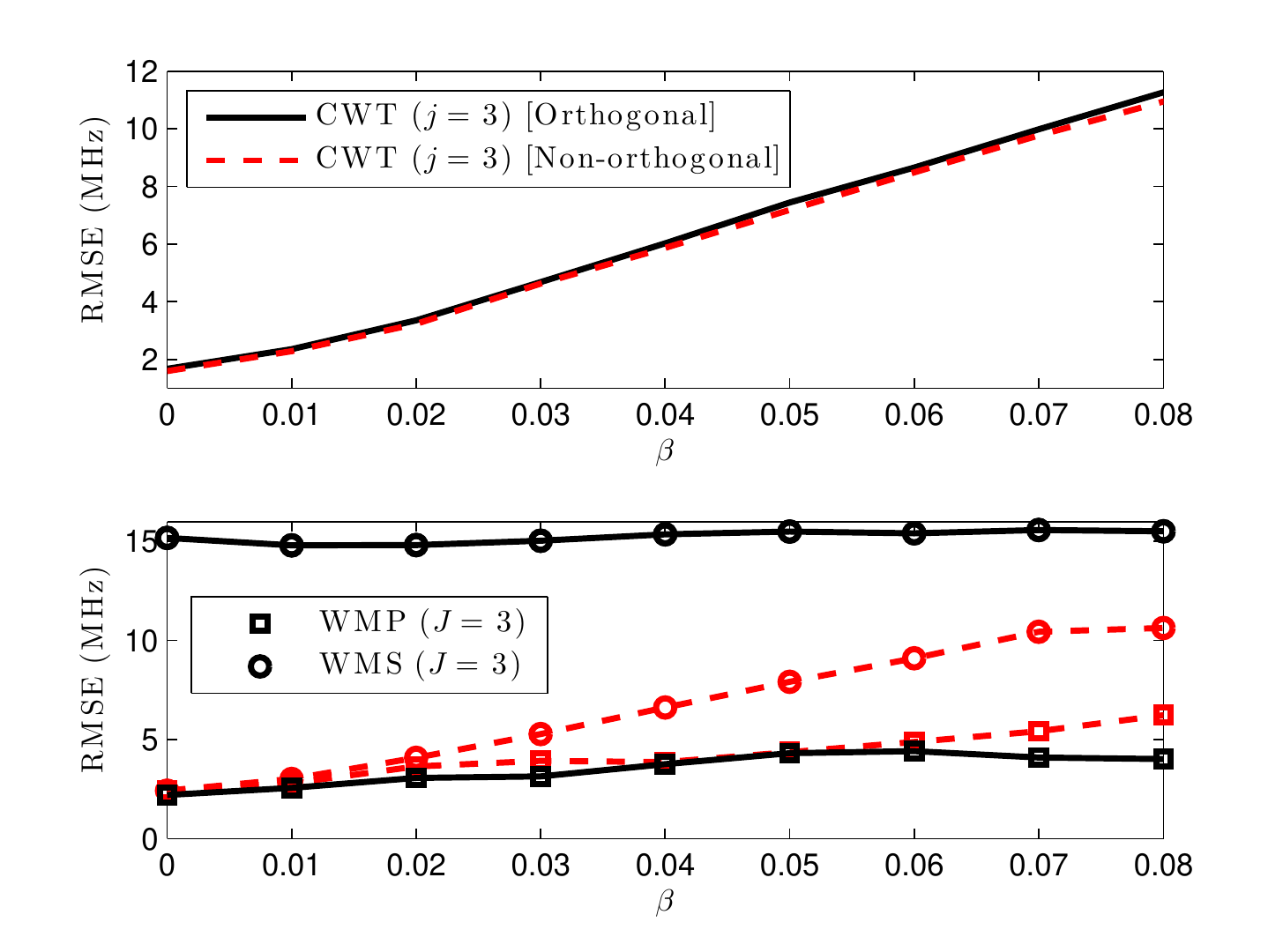}
\caption{The RMSE with variations of $\beta$.}
\label{fig:RMSE}
\end{figure}

Finally, the existing analysis on the wavelet-based techniques have focused on detecting edges of an ideal spectrum, where the transition from an idle subchannel to a busy one is sharp. However, these changes are not abrupt in practice, but rather they tend to be smooth. To understand how this smooth transition affects the edge-detection, we feed the PSD in Fig. \ref{fig:CWT} to a raised-cosine filter with a roll-off factor of $\beta$. Fig. \ref{fig:RMSE} illustrates the root-mean-square error (RMSE), $\sqrt{\mathbb{E}[(\mathbf{f}-\mathbf{\hat{f}})^2]}$, where $\mathbf{f}$ and $\mathbf{\hat{f}}$ are the true and estimated frequency boundaries, respectively, and $\mathbb{E}[.]$ is the expectation operator. We compare between the CWT, the WMP, and the WMS using two different families: The first-order Daubechies (orthogonal), and the Gaussian (non-orthogonal) wavelets. The RMSE of 1000 trials is computed. It it is clear that a slight increase of $\beta$ tangibly increases the RMSE of the CWT, whereas the WMP and WMS are more robust especially for the orthogonal wavelet. We also observe that the WMP has lower RMSE than the WMS. Also, the WMS performs better when a non-orthogonal wavelet is used instead of an orthogonal one.

In summary, further advancements are required to provide a robust algorithm that successfully detects subchannel edges and neglects false edges at low complexity. Also, different smoothing functions should be studied to analyze their impact on the quality of edge detection. Perhaps an adaptive algorithm that implements both WMS and WMP or both orthogonal and non-orthogonal smoothing functions can be further explored.

\section{Compressive Spectrum Sensing}

\subsection{The Compressive Sensing Problem}
One of the most celebrated theorems in signal processing is the Nyquist-Shannon Theorem, which states that to successfully reconstruct the sampled signal, the sampling rate must be at least as twice the maximum frequency component in that signal (also known as the Nyquist rate). In practice, almost all communication systems use sampling rate that is even higher than the Nyquist rate. Therefore, if we assume an SU with conventional signal processing to detect a wideband spectrum (say 1GHz), then the sampling rate must be at least 2GHz. This requirement demands unaffordable analog-to-digital (A/D) converters, which makes real implementations infeasible.

The literature on signal processing has been recently oriented towards techniques where sampling can be done below the Nyquist rate. This can be enabled by compressive sampling, and thus spectrum sensing done using this technique is referred as compressive sensing (CS).
Let $\mathbf{\hat{r}}=[\mathbf{r}_1^T,\mathbf{r}_2^T,\ldots,\mathbf{r}_k^T]^T$ be an $L(=KN)\times1$ vector that represents the wideband spectrum. Then, we can express it as
\begin{equation}
\label{eq:CompressiveSensing1}
\mathbf{\hat{r}}  =   \mathbf{B} \mathbf{y},
\end{equation}
where $\mathbf{B}$ is an $L \times L$ \emph{basis matrix}, and $\mathbf{y}$ is an $L\times1$ weighting vector.
There are two important definitions:
\begin{itemize}
\item We say that $\mathbf{r}$ is \emph{compressible} if the number of large coefficients in $\mathbf{y}$ is small.
\item We say that $\mathbf{r}$ is $Y$-sparse if it is a linear combination of only $Y$ basis vectors (i.e., $Y$ coefficients of $\mathbf{y}$ are non-zero and the rest are zero).
\end{itemize}

Assume that the SU receiver makes $M$ measurements such that $M<<L$, then the compressive sensing problem is described by \cite{Baraniuk}
\begin{equation}
\label{eq:CompressiveSensing2}
\mathbf{m} =   \boldsymbol{\Theta} \mathbf{\hat{r}}    =  \boldsymbol{\Theta} \mathbf{B} \mathbf{y},
\end{equation}
where $\mathbf{m}$ is an $M\times1$ \emph{measurement vector}, and $\boldsymbol{\Theta}$ is an $M \times L$ \emph{measurement matrix}. The CS problem is
\begin{itemize}
\item Design a stable $\boldsymbol{\Theta}$ such that we reduce the dimension of $\mathbf{\hat{r}}\in \mathbb{R}^L$ to $\mathbf{m}\in \mathbb{R}^M$ without incurring tangible signal information loss. It is shown in \cite{Candes}, that this matrix must be incoherent (i.e., the rows of $\boldsymbol{\Theta}$ cannot sparsely represent the columns of $\mathbf{B}$).
\item Reconstruct $\mathbf{\hat{r}}$ from only $M$ measurements of $\mathbf{m}$ instead of $L$ samples. This requires advanced reconstructions algorithms.
\end{itemize}

\subsection{Challenges}
There are some challenges that need to be further explored when CS is implemented. Signal sparsity is an important requirement to tangibly reduce the sampling rate of the A/D. Recall that one of the main motivations of the introduction of cognitive radio (CR) is that the frequency spectrum is underutilized (i.e., sparse). Thus, most of the compressive sensing techniques exploit sparsity in frequency domain  \cite{Tian2, Tian1, Tian4, Tian5}. However, upon the implementation of CR, the spectrum utilization should be enhanced, which means that the spectrum loses its sparsity in frequency. In other words, CS is a solution for today's problem, and this solution will become a problem for CS in the future! Therefore, other domains must be investigated. For example, the cyclostationarity detector exploits features from the \emph{cyclic} spectrum. This spectrum is shown to be sparse since the PUs do not exploit all cyclic frequencies \cite{Tian3}.

Another issue is the expected degradation of the SNR since we are only using a partial number of measurements. In fact, one of the key issues is the number of measurements required, $M$, which directly depends on the sparsity level. Therefore, either a good estimation of the sparsity level is required, or else an adaptive number of measurements is needed when the sparsity level is unknown or not fixed.

While in conventional sampling, the reconstructed signal is simply a linear combination of the received measurements, in compressive sampling, the reconstruction procedure is nonlinear \cite{Tropp}. In other words, we are trading the hardware burden with software burden. Another issue is the design of sparsity basis. It is usually assumed that this is known at the SU side; however, a more robust algorithm is required when it is unknown. This is even more challenging when there are multiple SUs because synchronization of such bases among SUs becomes essential.

\section{Angle-based Spectrum Sensing}
\begin{figure}[!t]
\centering
\includegraphics[width=3.5in]{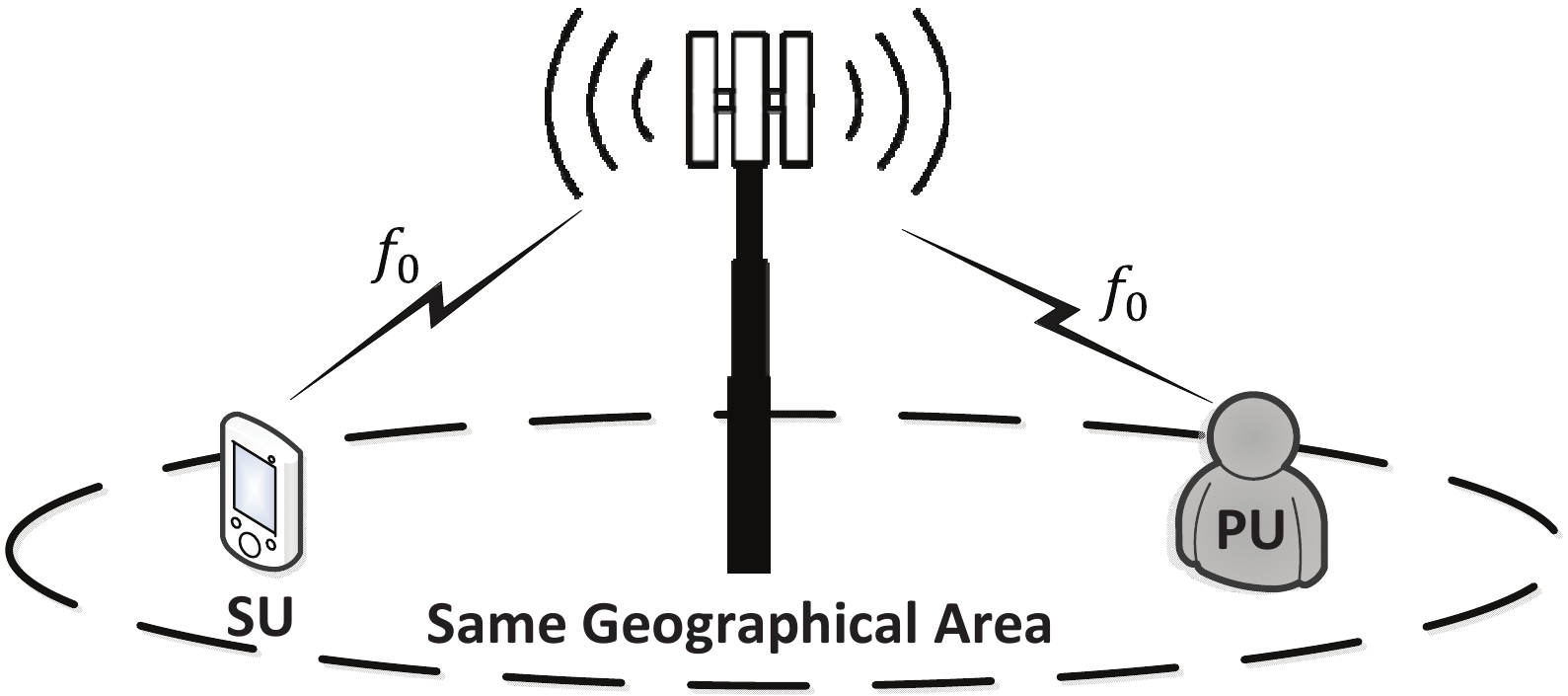}
\caption{The angle-domain provides additional spectral opportunities.}
\label{fig:Angle}
\end{figure}

Due to the advancements in multi-antenna technologies such as beamforming, array processing and multiple-input multiple-output (MIMO) systems, opportunities in the angle-domain can be utilized for spectrum sharing in addition to the conventional domains: time, frequency and space. The basic principle is that the SU estimates the direction of arrival (DOA) of the PU signal. Once the azimuth angles are determined, the SU can estimate the PU's position \cite{Mahram}. Then, the SU can transmit at the same time with the PU, in the same area, and at the same frequency as long as it steers its transmission over a different angle as illustrated in Fig. \ref{fig:Angle}. The main issue here is that the SU receiver must be equipped with multiple antennas, and this will incur additional hardware and software complexities.

In addition, using a multi-antenna array, the SU can divide the antennas into two subsets: sensing antennas (SA) and transmission antennas (TA). Thus, the SU can potentially sense and transmit at the same time. This significantly enhances the throughput of the CRN. For instance, if the DOA of the PU signal is estimated, the SU's TA can be steered towards an orthogonal direction with respect to the position of the PU. Another example is the \emph{TranSensing} architecture proposed in \cite{Daesik}. In TranSensing, the SA determine if the PU is present or not. At the same time, the SU can transmit its data via the TA. This, however, requires isolating the antennas with reliable echo interference cancellation techniques \cite{Daesik} (the reader may refer to \cite{Tsakalaki} for such techniques).

\section{Conclusion}
Multiband spectrum access has many great potentials for CRNs. However, many challenges and issues must be overcome to efficiently exploit these potentials. In this paper, we discuss some of these challenges. Particularly, we have shown that further advancements in wavelet-based techniques are required to mitigate some of the issues such as: the presence of false-edges, the rejection of true-edges, and dealing with a realistic spectrum (where the edges are smooth). Moreover, some of challenges in compressive sensing are presented (e.g., unknown basis matrix, sparsity in different domains, etc.). Finally, we have discussed the potential of the angle-domain, and we argue that additional spectral opportunities can be utilized if reliable angle-based sensing algorithms are implemented.

\bibliographystyle{IEEEtran}
\bibliography{C:/Users/Ghaith_90/Dropbox/Latex/IEEEabrv,C:/Users/Ghaith_90/Dropbox/Latex/References}

\end{document}